\documentclass[aps,prl,twocolumn,reprint]{revtex4-1}

\usepackage{amsmath,hyperref,graphicx,color,tikz}
\usepackage{ulem,soul}
\usepackage{graphicx}
\usepackage{subfigure}
\usepackage{verbatim}
\usepackage{varwidth,xcolor}
\usepackage{booktabs}

\hypersetup{
    colorlinks=true,       
    linkcolor=blue,          
    citecolor=blue,        
    filecolor=blue,      
    urlcolor=blue           
}

\newsavebox{\graphicsbox}

\graphicspath{{./}{figures/}}

\begin{document}

\title{Temperature induced decay of persistent currents in a superfluid ultracold gas} 

\author{A. Kumar}
\author{S. Eckel}
\author{F. Jendrzejewski}
\author{G. K. Campbell}
\email{gcampbe1@umd.edu}
\affiliation{Joint Quantum Institute, National Institute of Standards and Technology and University of Maryland, Gaithersburg, Maryland 20899, USA}

\date{\today}

\begin{abstract}  
We study how temperature affects the lifetime of a quantized, persistent current state in a toroidal Bose-Einstein condensate (BEC). When the temperature is increased, we find a decrease in the persistent current lifetime. Comparing our measured decay rates to simple models of thermal activation and quantum tunneling, we do not find agreement. We also measured the size of hysteresis loops size in our superfluid ring as a function of temperature, enabling us to extract the critical velocity. The measured critical velocity is found to depend strongly on temperature, approaching the zero temperature mean-field solution as the temperature is decreased. This indicates that an appropriate definition of critical velocity must incorporate the role of thermal fluctuations, something not explicitly contained in traditional theories. 
\end{abstract}


\maketitle
Persistent currents invoke immense interest due to their long lifetimes, and they exist in a number of diverse systems, such as  superconductors~\cite{onnes1914,file1963}, liquid helium~\cite{mehl1968,rudnick1969}, dilute ultracold gases~\cite{ryu2007,ramanathan11,beattie13} and polariton condensates~\cite{sanvitto10}. Superconductors in a multiply connected geometry exhibit quantization of magnetic flux,~\cite{doll1961} while the persistent current states in a superfluid are quantized in units of $\hbar$, the reduced Planck constant. To create transitions between quantized persistent current states, the critical velocity of a superfluid (or critical current of a superconductor) must be exceeded.  In ultra-cold gases, the critical velocity is typically computed at zero-temperature, whereas experiments are obviously performed at non-zero temperature. In this work, we experimentally investigate the role of temperature in the decay of persistent currents in ultracold-atomic, superfluid rings (Fig. 1a).

\begin{figure}
\centering
\includegraphics[width=80mm]{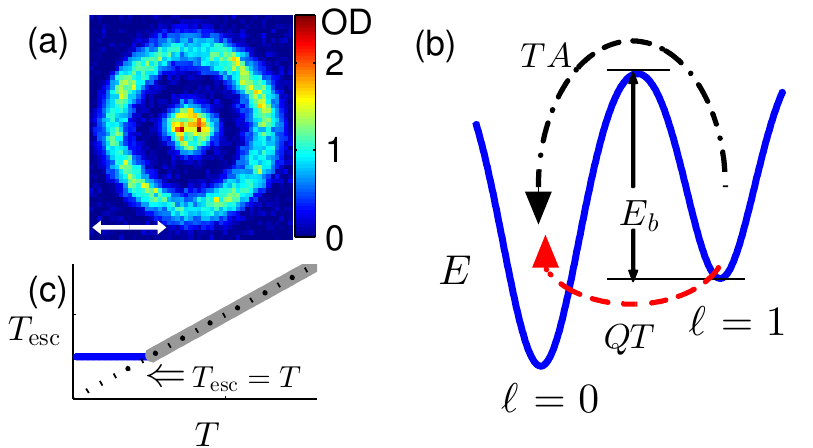}     
\caption{\label{fig:decayschematic} Target shaped condensate, energy landscape  and effectice escape temperature (color online). a) {\it In situ} image of trapped atoms, with 5\% of the total atoms imaged \cite{ramanathan2012rsi}. Experiments are performed on the ring-shaped BEC and the resulting winding number $\ell$ is read out by interfering the ring condensate with the disc-shaped BEC in time of flight. The disc-shaped BEC acts as a phase reference. (b) Energy landscape showing the stationary state, $\ell=0$, and the persistent current state, $ \ell =1$, as minima in the potential. The energy barrier $E_{b}$ needs to be overcome for a persistent current to decay from $\ell=1$ to $\ell=0$. The decay can be induced either via thermal activation (TA), or quantum tunneling (QT). (c) Crossover from quantum tunneling to the thermally activated regime. The escape temperatre $T_{\rm esc}$ (see text) first remains constant (horizontal blue line) and the becomes equal to the physical temperature $T$(slanted gray line). A dotted line acts a guide to the eye depicting $T_{\rm esc}=T$.}
\end{figure}

In the context of the free energy of the system, different persistent current states of the system (denoted by an integer $\ell$ called the winding number) can be described by local energy minima, separated by energy barriers (here, we concentrate on $\ell=0$ and $\ell=1$ shown in Fig.1(b))~\cite{mueller2002,eckel2014h}. The metastable behavior emerges from the energy barrier, $E_{\rm b}$, between two persistent current states. For superconducting rings, the decay dynamics have been understood by the Caldeira-Leggett model~\cite{caldeira1983}: the decay occurs either via quantum tunneling through the energy barrier or thermal activation over the top of the barrier. When first investigated in superconductors~\cite{martinis1985,martinis1988,martinis1988b,rouse1995}, the decay rate from the metastable state $\Gamma$ was fit to an escape temperature $T_{\rm esc}$ by the relation $\Gamma = \Omega_a \exp(E_{\rm b}/k_{B}T_{\rm esc})$, where $k_{\rm B}$ is the Boltzmann constant. In the context of the WKB approximation in quantum mechanics or the Arrhenius equation in thermodynamics, $\Omega_a$ represents the ``attempt frequency'': i.e. how often the system attempts to overcome the barrier. The $\exp(E_b/k_B T_{esc})$ represents the probability of surmounting the barrier on any given attempt. The probability and thus the escape temperature in quantum tunneling is independent of temperature, while for thermal activation, the escape temperature tracks the real temperature (Fig \ref{fig:decayschematic}(c)). For our superfluid ring, the energy barrier $E_b$ is much greater than all other energy scales in the problem, hence the lifetime of the persistent current is much greater than the experimental time-scale. However, the height of the energy barrier and the relative depth of the two wells can be changed by the addition of a density perturbation~\cite{eckel2014h}. The density perturbation may induce a persistent current decay even if its strength is less than the chemical potential~\cite{ramanathan11,eckel2014h}.

In this paper, we measure the decay constant of a persistent current for various perturbation strengths and temperatures. We also measure the size of hysteresis loops which allows us to extract the critical velocity, showing a clear effect of temperature on the critical velocity in a superfluid.

\begin{figure}
\centering
\includegraphics[width=80mm]{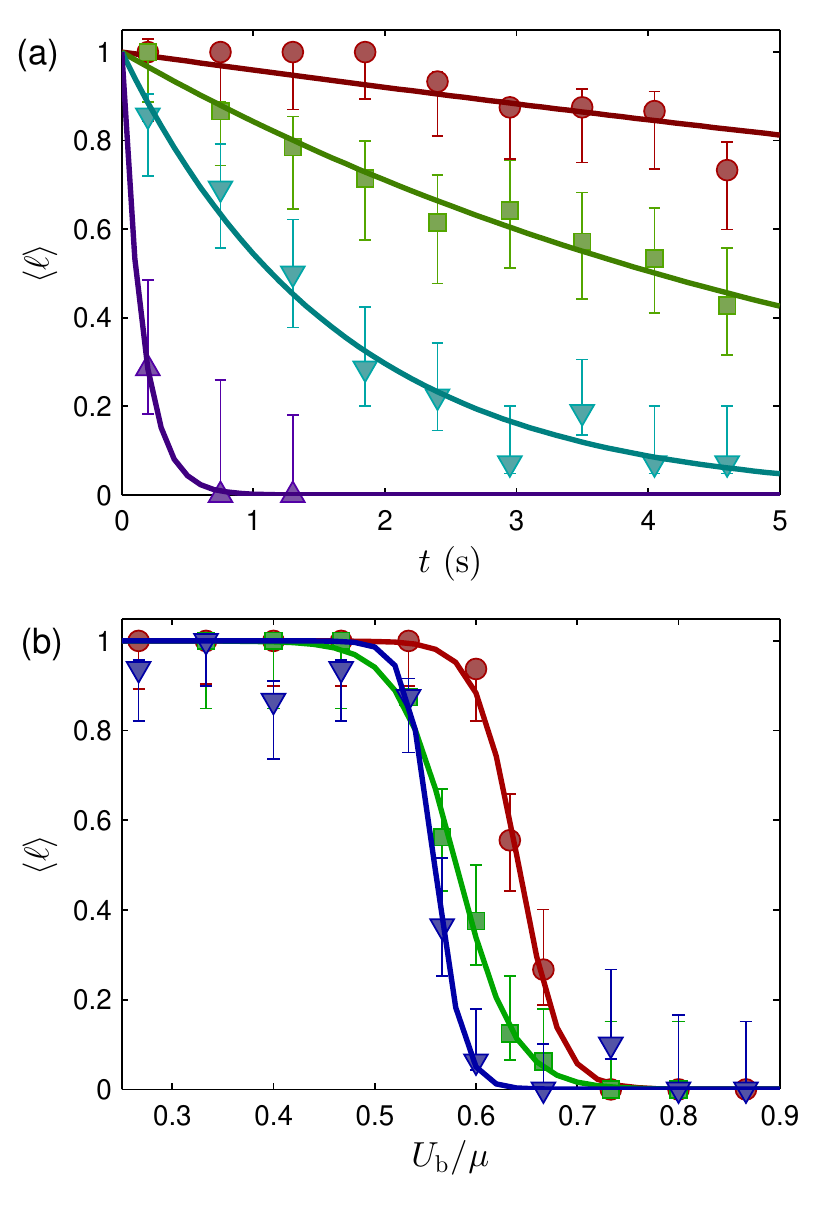}    \caption{\label{fig:circdecay} (color online). (a) Average measured winding number $\langle \ell \rangle$ vs. $t$, the duration for which a stationary perturbation is applied. The four data sets correspond to different strengths of the stationary perturbation $U_b$: 0.50(5)$\mu$ (circles), 0.53(5)$\mu$ (squares), 0.56(6)$\mu$ (inverted triangles) and 0.59(6)$\mu$ (triangles). Here, $\mu$ is the unperturbed chemical potential. The temperature of the superfluid was 85(20)~nK. The solid curves show exponential fits. (b) The average measured winding number $\langle \ell \rangle$ vs. $U_b$ for fixed $t$: 0.5~s (circles), 2.5~s (squares) and 4.5~s (inverted triangles). The solid curves show a sigmoidal fit of the form  $\langle \ell \rangle = [\exp(({U_{b}}/ \mu -\zeta)/\alpha)+1]^{-1}$. The temperature of the superfluid was 40(12)~nK.}
\end{figure}

The preferred theoretical tool for modeling atomic condensates is the Gross-Pitaevskii (GP) equation, which is a zero-temperature, mean-field theory. Recent experiments exploring the effect of rotating perturbations on the critical velocity of toroidal superfluids have found both agreement~\cite{fred2014,ryu2013} and significant discrepancies~\cite{eckel2014h,ramanathan11} between experimental results and GP calculations. Several non-zero temperature extensions to GP theory have been developed, including ZNG~\cite{zaremba1999} and c-field~\cite{blakie2008} [of which the Truncated Wigner approximation (TWA) is a special type]. To explore the role of temperature in phase slips in superfluid rings, Ref.~\cite{ludwig2014} studied condensates confined to a periodic channel using TWA simulations. In addition, recent theoretical~\cite{rooney2010,rooney2016,kobayashi2006,jackson2009,duine04,berloff2007,fedichev99} and experimental~\cite{shin2015b} works explored a similar problem of dissipative vortex dynamics in a simply-connected trap.

Our experiment consists of a $^{23}$Na Bose-Einstein condensate (BEC) in a target-shaped optical dipole trap~\cite{eckel2014cp} [Fig. \ref{fig:decayschematic}(a)]. The inner disc BEC has a measured Thomas-Fermi (TF) radius of 7.9(1)~$\mu$m. The outer toroid has a Thomas-Fermi full-width of 5.4(1)~$\mu$m and a mean radius of 22.4(6)~$\mu$m. To create the target potential, we image the pattern programmed on a digital micromirror device (DMD) onto the atoms while illuminating it with blue-detuned light. This allows us to create arbitrary potentials for the atoms. Vertical confinement is created either using a red-detuned TEM$_{00}$ or a blue-detuned TEM$_{01}$ beam. The potential generated by the combination of the red-detuned TEM$_{00}$ beam and ring beam is deeper than that of blue-detuned TEM$_{01}$ and ring beam; thus the temperature is generally higher in the red-detuned sheet potential.   We use this feature to realize four different trapping configurations with temperatures $T$ of 30(10)~nK, 40(12)~nK, 85(20)~nK and 195(30)~nK but all with roughly the same chemical potential of $\mu/\hbar= 2\pi\times(2.7(2)\mbox{ kHz})$.  (See supplemental material for details about temperature and trapping configurations.) Finally, a density perturbation is created by another blue-detuned Gaussian beam with a $1/e^2$ width of 6~$\mu$m and can be rotated or held stationary at an arbitrary angle in the plane of the trap~\cite{kevin2013}.

To probe the lifetime of the persistent current, we first initialize the ring-shaped BEC into the $\ell=1$ state with a fidelty of 0.96(2) (see Supplemental material). A stationary perturbation with a strength $U_{b}<\mu$ is then applied for a variable time $t$ ranging from 0.2~s to 4.6~s. To compensate for the $25(2)$~s lifetime of the condensate, we insert a variable length delay between the initialization step and application of the perturbation to keep the total time constant (Without this normalization, a $25(2)$~s lifetime would cause an atom loss of $\approx$ 20~\% in 4.7~s, changing the chemical potential by $\approx 10$~\%). At the end of the experiment, the circulation state is measured by releasing the atoms and looking at the resulting interference pattern between the ring and disc BECs~\cite{eckel2014h,corman2014}. For each temperature, four different perturbation strengths are selected.  The perturbation strengths are chosen such that the lifetime of the persistent current state is varied over the entire range of $t$. The measurement is repeated 16-18 times for each combination of $U_b$, $T$ and $t$. The average of the measured circulation states $\langle \ell \rangle$ gives the probability of the circulation state surviving for a given set of experimental parameters.

Figure \ref{fig:circdecay}(a) shows $\langle \ell \rangle$ vs. $t$ for $T=85(20)$~nK and four different $U_b$. We fit the data to an exponential $\exp(-\Gamma t)$. GP theory predicts either a fast decay ($<10$~ms) or no decay, depending on the precise value of $U_b/\mu$ \cite{ludwig2014}. By contrast, we see from Fig.~\ref{fig:circdecay}(a) that $\Gamma$ changes smoothly from $4.1(6)\times10^{-2}$~s$^{-1}$ to 6.2(8)~s$^{-1}$ as $U_b$ is changed from $0.50(4)\mu$ to $0.59(5)\mu$. Thus we are able to tune the decay rate by over two orders of magnitude by changing the magnitude of perturbation by $\approx 0.1\mu$, in qualitative agreement with TWA simulation results~\cite{ludwig2014}. This confirms that the decay of a persistent current is a probabilistic process, in contrast to the instananeous, deterministic transitions seen in GPE simuations~\cite{ludwig2014}.

To explore whether a longer hold time shifts or broadens the transition between persistent current states, we measured the average persistent current as a function of $U_b$ while keeping $t$ constant. Figure~\ref{fig:circdecay}(b) shows this measurement for three different $t$: 0.5~s, 2.5~s and 4.5~s. We fit this data to a sigmoidal function of the form $\langle \ell \rangle = [\exp(({U_{b}}/ \mu -\zeta)/\alpha)+1]^{-1}$ to extract estimates of the width $\alpha$ and center $\zeta$ of the transition~\footnote{the extracted center and FWHM of the transition are independent of the form of the sigmoidal funcation chosen}. We see that changing the perturbation strength by $\approx 0.2 \mu$ decreases $\langle \ell \rangle$ from one to zero. The width $\alpha$ is essentially unchanged as we change $t$ from 0.5~s to 4.5~s, though the center of the sigmoid $\zeta$ shifts by $\approx 0.1 U_b/\mu$. We also took similar measurements at a temperature of 85(20)~nK (not shown). The width $\alpha$ remains essentially independent of $t$ even at higher temperatures. For a hold time $t=0.5$~s, we found a center $\zeta=0.50(4)U_b/\mu$ at $T=85(20)$~nK; by contrast, we obtain $\zeta=0.64(4)U_b/\mu$ for a $T=40(12)$~nK. This indicates that an increase in temperature makes a phase slip more probable even with smaller $U_b$.

\begin{figure}
\centering
\includegraphics[width=80mm]{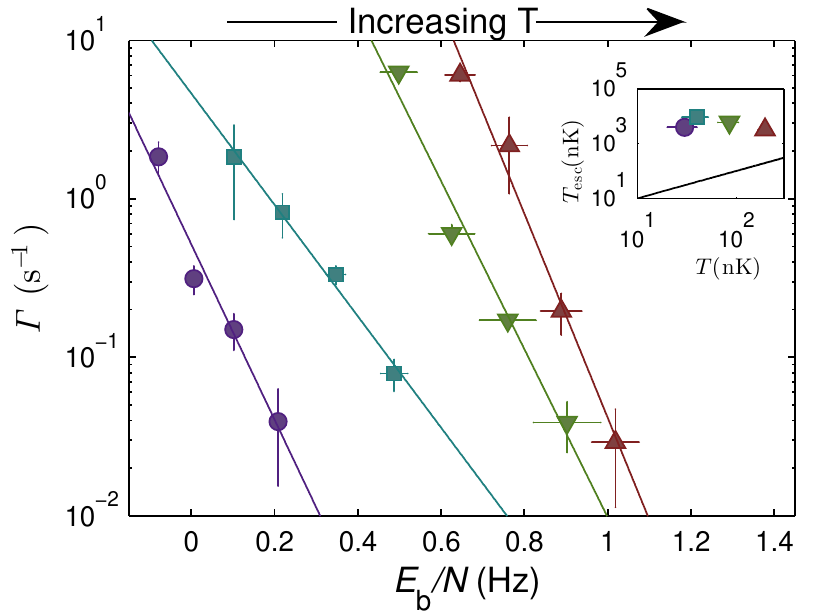}
\caption{\label{fig:logdecay}  (color online). Measured decay rate of the persistent current $\Gamma$ as a function of perturbation strength $U_b$ for four different temperatures: 30(10)~nK (circles), 40(12)~nK (squares), 85(20)~nK (inverted circles) and 195(30)~nK (triangles). The solid lines are fits of the form $\Gamma = \Omega_a  \exp(E_b/k_B T_\text{esc} ) $, where $E_b$ is the energy barrier, $k_{B}$ is the Boltzmnann constant, and $T_\text{esc}$ and $\Omega_a$ are fit parameters. The inset shows the extracted $T_\text{esc}$ as a function of measured physical temperature: 30(10)~nK (triangle), 40(12)~nK (square), 85(20)~nK (circle) and 195(30)~nK (inverted triangle). The solid line shows $T_{\text esc}=T$.}
\end{figure}

To understand if the decay of the persistent current is thermally activated or quantum mechanical in nature, we first must understand the nature of the energy barrier, $E_b$, that separates the two states.  To estimate the size of $E_b$, we consider excitations that connect the $\ell=1$ to the $\ell=0$ state. In the context of a one-dimensional ring, a persistent current decay corresponds to having either thermal or quantum fluctuations reduce the local density, producing a soliton that subsequently causes a phase slip~\cite{little1967}.  For rings with non-negligible radial extent, TWA simulations suggest that a vortex passing through the annulus of the ring (through the perturbation region) causes the transition~\cite{ludwig2014}. Because of the narrow width of our ring, we expect that a solitonic-vortex is the lowest energy excitation that can connect two persistent current states~\cite{brand2001,brand2015,zwierlein2014,valtolina1505,ferrari2014,ferrari2015}. An analytical form for the energy of a solitonic vortex is given by~\cite{zwierlein2014,brand2015}:
\begin{equation}
\epsilon_{sv} (U_{ b}/\mu) \approx \pi n_{2D} \frac{\hbar^2}{m} \ln (\frac{R_{\perp}}{\xi}) + \frac{1}{2} m N_c \left ( \frac{\hbar}{2 mR} \right ) ^2
\end{equation}
where $N_c$ is the total number of condensate atoms in the ring, $\xi$ is the healing length, $R_{\perp}$ is the Thomas-Fermi width of the perturbation region and $n_{2D}$ is the maximum 2D density in the region of the perturbation. The first term is the energy of a solitonic-vortex while the second term is the kinetic energy of the remaining $\pi$ phase winding around the ring. We note that $N_c$, $R_{\perp}$, $\xi$ and $n_{2D}$ all depend implicitly on $T$ and $U_b$. Finally,
\begin{equation}
 E_b(U_b,T) = \epsilon_{sv} - \epsilon_{\ell=1} = \epsilon_{sv} - \frac{1}{2} m N_c\left(\frac{\hbar}{mR}\right)^2,
\label{eq:energyfunc}
\end{equation}
where $\epsilon_{\ell=1}$ is the energy of the first persistent current state. We have verified the accuracy of these expressions using GP calculations similar to those in Refs.~\cite{brand2002,komineas2003,zwierlein2014,brand2015} to within 10~\% for our parameters.   

Fig.~\ref{fig:logdecay} shows the clear temperature dependence of the measured decay rate $\Gamma$ of the persistent current. To quantify this dependence, we fit the data to the form $\Gamma = \Omega_a \exp (E_b/k T_\text{esc})$ for each temperature (shown as the solid lines in Fig.~\ref{fig:logdecay}). We note that while the attempt frequency $\Omega_a$ is dependent on temperature (changing by five orders of magnitude from 40(12)~nK to 195(30)~nK), $T_\text{esc}$ is not (see inset of Fig.~\ref{fig:logdecay}). In fact, $T_{\rm esc}$ is roughly constant at $\approx 3\mu K$, while the BEC temperature varies from 30(10)~nK to 195(30)~nK. Thus, simple thermal activation does not explain the probability of a transition, since $T_\text{esc} \neq T$. The constancy of $T_\text{esc}$ hints that a temperature-independent phenomenon like macroscopic quantum tunneling may play a role, as it does in superconducting systems~\cite{voss1981}.  We can estimate the decay rate due to quantum tunneling by drawing an analogy with an rf-superconducting quantum interference device. In this device, the quantum tunneling rate can be estimated by the WKB approximation, $\Gamma \approx (\omega_p/2\pi)\exp(-E_b/\hbar\omega_p)$, where $\omega_p$ is the frequency of the first photon mode in the superconducting system~\cite{martinis1988}. Here, by analogy, $\omega_p$ is the frequency of the first azimuthal phonon mode, which is $\approx 2\pi\times30$~Hz. For our system, $E_b/\hbar\omega_p>10^3$, so the quantum tunneling should be negligible.  Thus, the observed decay cannot cannot be described by either simple thermal activation or quantum mechanical tunneling.  It may be that more complicated models of energy dissipation may be required.  

\begin{figure}
\centering
\includegraphics[width=80mm]{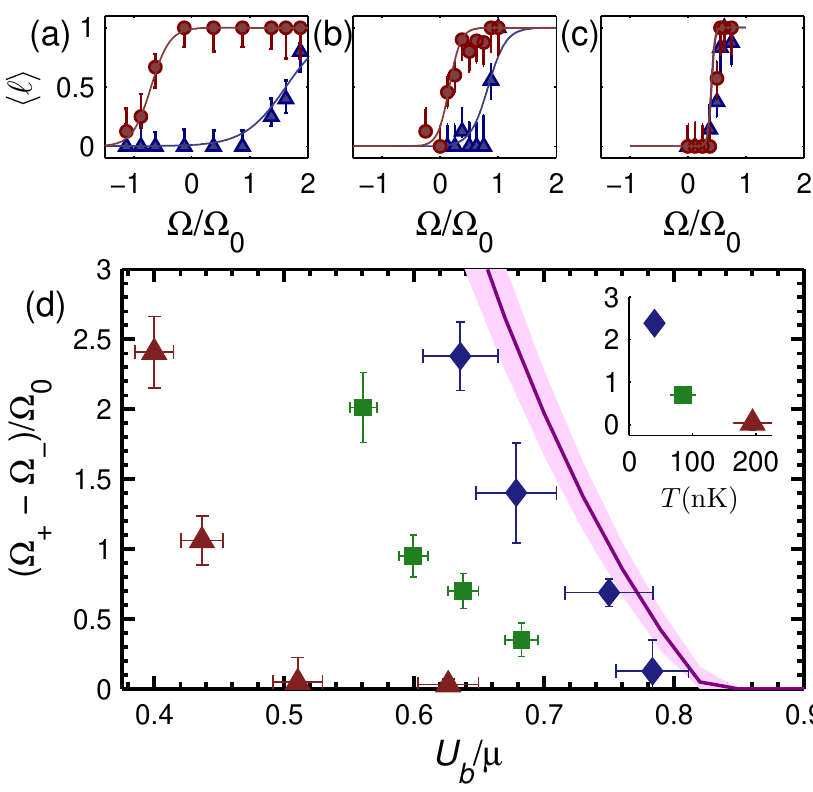}  
\caption{\label{fig:hysteresissummary} 
Hysteresis loop for a perturbation strength of $0.64(4) U_b/\mu$ for 40(12)~nK (a), 85(20)~nK (b), and 195(30)~nK (c).(d) Size of the hysteresis loop, $(\Omega_+-\Omega_-)/\Omega_0$ (see text), vs. barrier strength for three different temperatures: 40(12)~nK, diamonds, 85(12)~nK (squares), and 195(12)~nK (triangles). The zero temperature, GPE predicted, area of the hysteresis loop is shown as a purple band, which incorporates the uncertainty in speed of sound. The inset shows the hystersis loop size shown in (a)-(c) as a function of temperature for a perturbation strength of $0.64(4) U_b/\mu$.}
\end{figure}

Finally, because there are parallels between a vortex moving through the annulus of the ring and a vortex leaving a simply connected BEC, we investigated models that predict the dissipative dynamics of these vortices~\cite{fedichev99,duine04}. Such models predict lifetimes that scale algebraically with $E_b$ and $T$. As can be seen from Fig.~\ref{fig:logdecay} our data scales exponentially with $E_b$. Thus, these models fail to explain the experimental data.

The measurements of the decay constants described above shows the strong effect  of temperature on the persistent current state. As discussed above, this temperature dependence is wholly captured in the variation of the constant $\Omega_a$ with $T$, as $T_\text{esc}$ is constant. This causes an apparent change in the critical velocity of a moving barrier (for a given application time), with higher temperatures having lower critical velocities. Such a change in critical velocity affects hysteresis loops \cite{eckel2014h}. For initial circulation state $\ell=0$(1), we experimentally determine $\Omega_+$($\Omega_-$), the angular velocity of the perturbation at which $\langle \ell \rangle=0.5$. The hysteresis loop size is given by $\Omega_{+}-\Omega_{-}$, normalized to $\Omega_0$, where $\Omega_0=\hbar/mR^2$, $m$ is the mass of an atom, $R$ is the mean radius of the torus. We measure the hysteresis loop for four perturbation strengths and three different temperatures: 40(10)~nK, 85(20)~nK and 195(30)~nK as shown in Fig.~\ref{fig:hysteresissummary} , with the zero-temperature GP prediction based on the speed of sound shown for references~\cite{eckel2014h,watanabe09}. We see from Fig.~\ref{fig:hysteresissummary} that the discrepancy between experimental data and theoretical predictions decreases as the temperature is lowered. Using the density distribution of atoms around the ring, we extract the critical velocity from the hysteresis loop size~\cite{eckel2014h}. For example, at $U_b/\mu = 0.64(4)$, a temperature change of 40(12)~nK to 195(30)~nK corresponds to a change in the critical velocity of 0.26(6)~$c_{\rm s}$ to 0.07(3)~$c_{\rm s}$. Here, $c_{\rm s}$ is the speed of sound in the bulk. While the measured critical velocity approached the zero-temperature, speed of sound, we see that at non-zero temperature thermal fluctuations must be taken into account in any measurement or calculation of the critical velocity.

In conclusion, we have measured the effect of temperature on transitions between persistent current states in a ring condensate in the presence of a local perturbation. The results of this work indicate that as thermal fluctuations become more pronounced, it becomes easier for the superfluid to overcome the energy barrier and the persistent current state to decay. If we assume that the decay is thermally driven and is thus described by an Arrhenius-type equation, we find a significant discrepancy between the measured temperature and the effective temperature governing the decay. Other possible mechanisms like macroscopic quantum tunneling should be greatly suppressed. Despite the disagreement, we find a clear temperature dependence of the critical velocity of the superfluid by measuring hysteresis loops. This work will provide a benchmark for finite temperature calculations on the decay of topological excitation in toroidal superfluids.

\begin{acknowledgments}
The authors thank M. Edwards, M. Davis, A. Yakimenko, and W.D. Phillips for useful discussions. This work was partially supported by ONR, the ARO atomtronics MURI, and the NSF through the PFC at the JQI.
\end{acknowledgments}

\clearpage
\newpage
\newpage

\section{Supplementary material for ``Temperature induced decay of persistent currents in a superfluid ultracold gas"}
This supplemental material contains three sections. The first section explains the experimental procedure for initializing the persistent current state and the subsequent measurement. The second section explains the procedure for extracting temperature. The third section presents the method we use to calibrate perturbation strength and the effect of finite temperature on the calibration of the perturbation strength.

\subsection{Experimental procedure}
\begin{figure}
\centering
\includegraphics[width=80mm]{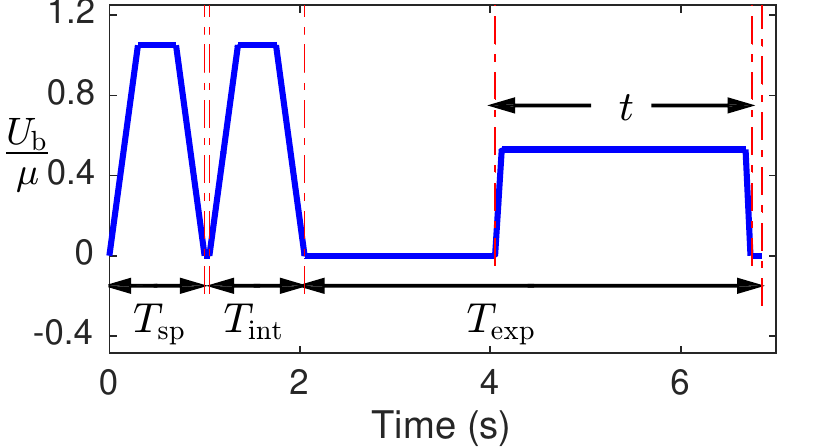}     
\caption{\label{fig:aodschematic} The experimental sequence for measuring current decay. A stationary perturbation with height $U_{\rm b}/\mu\approx1.1$, where $\mu$ is the chemical potential is turned on during $T_{\rm sp}$ to destroy any spontaneous circulation. A rotating perturbation with the same height imprints the $\ell=1$ circulation state during $T_{\rm int}$. A stationary perturbation with strength less than the chemical potential (shown here as 0.5$U_b/\mu$) probes the circulation state for $t$. An intermediate step $T_{\rm exp}-t$ is introduced to ensure that the total experimental time $T_{\rm exp}$ remains constant.}
\end{figure}

After creating a BEC in the target shaped trap, the experiment involves two stages, first a preparation stage followed by a measurement stage [see Fig.~\ref{fig:aodschematic}]. In the preparation stage, a stationary perturbation is adiabatically raised in the ring for a total time, $T_{\rm sp}=1$~s to destroy any spontaneous circulation states. Subsequently, a circulation state is imprinted on the atoms by moving the perturbation around the ring for a total time, $T_{\rm int}=1$~s. In the preparation stage, both perturbations have a strength of $\approx 1.1\mu$, where $\mu$ is the unperturbed chemical potential. The density perturbation is raised to this strength in 300~ms, kept constant for 400~ms and then lowered down to zero in 300~ms. The reliability of the experimental data depends both on our ability to imprint circulation states deterministically and to eliminate spontaneous circulation states. The confidence level of having no spontaneous circulation before imprinting the circulation state is 0.99(1). The confidence level of imprinting a circulation state with one unit of circulation before the measurement stage is 0.96(2).

To measure the decay constant, we again apply a stationary perturbation whose strength is variable, but always less than the chemical potential. The perturbation is applied for a variable time $t$, during which it is raised to a desired strength in 70~ms, kept constant and then lowered down in 70~ms.

To measure the hysteresis loop size, we initialize the atoms in the ring in either a circulation state of $\ell=0$ or $\ell=1$. A rotating perturbation with a strength less than the chemical potential $\mu$ is then applied with a variable rotation rate to trace out the hysteresis loop~\cite{eckel2014h1}. The rotating perturbation is on for a total of 2~s, during which it is raised to the desired strength in 300~ms, kept constant, and then lowered in 300~ms.

\subsection{Measuring the temperature}

\begin{figure}
\centering
\includegraphics[width=80mm]{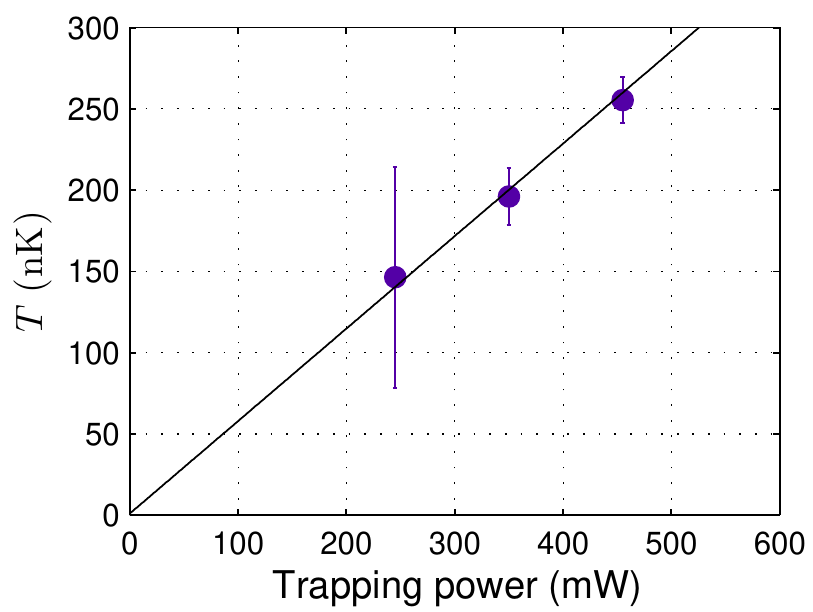}     
\caption{\label{fig:irsheettemp} Measured temperature vs. power in the red-detuned (1064~nm) vertical trapping beam. The errorbars show the statistical uncertainty.}
\end{figure}

The persistent current lifetime was measured at four different temperatures. The higher temperatures of 85(20)~nK and 195(30)~nK are achieved using the red-detuned vertical trap while the lower temperatures of 30(10)~nK and 40(12)~nK are obtained using the blue-detuned vertical trap. Typically, the temperature of the BEC is extracted by releasing the atoms from the trap and measuring the density distribution in time of flight (ToF). The 1D integrated density is then fitted to a bimodal distribution: the sum of a Gaussian and a Thomas-Fermi profile. The Gaussian part describes the thermal part while the Thomas-Fermi profile describes the condensate part. Fitting the evolution of the width of the Gaussian as a function of time yields the temperature~\cite{ketterle1999c}. 

To understand the final temperature, we need to understand the evaporation profile and the final trap configuration. During the evaporative cooling stage, the laser cooled atoms are transferred to a red-detuned optical dipole trap with a depth of the order of 10~$\mu$K. We then do an exponential forced evaporation ramp by lowering the laser power to obtain a degenerate quantum gas. The temperature of this gas is set by the final depth of the optical dipole trap. We reach a temperature of 85(20)~nK and 195(30)~nK for powers of 140~mW and 350~mW of red-detuned IR light respectively. A separate TEM$_{00}$ red-detuned crossed dipole trap is then turned on~\footnote{The transverse dimensions of atoms in the red-detuned vertical trapping beam is on the order of 100~$\mu$m, while the target trap is only $\approx$50~$\mu$m in diameter. To ensure efficient transfer to the target trap, a Gaussian beam of $1/e^2$ width of $\approx$ 50~$\mu$m is turned on in tandem with the vertical confinement beams (effective mode matching)}, after which the condensate is transferred to the target trap. The atoms now reside in a potential which is the convolution of an attractive potential of the red-detuned sheet trap and a repulsive blue-detuned target trap. The trap depth and hence the temperature is set by the red-detuned trap, since the potential due to red-detuned TEM$_{00}$ beams are typically deeper than their blue-detuned counterparts~\cite{friedman2002}. To extract the temperature, we release the atoms in the target trap in time of flight and then image the cloud in the horizontal direction. We extract a temperature by fitting the atom density to a bimodal distribution. This measurement was repeated at various optical powers. The temperature of 195(30)~nK at 350~mW of trap power can be measured directly. The temperature of 85(20)~nK at 140~mW is obtained by extrapolation of the fit shown in Fig.~\ref{fig:irsheettemp}. This extrapolation is necessary as the bimodal fit becomes less reliable at lower temperature, as the thermal fraction decreases.

\begin{figure}
\centering
\includegraphics[width=80mm]{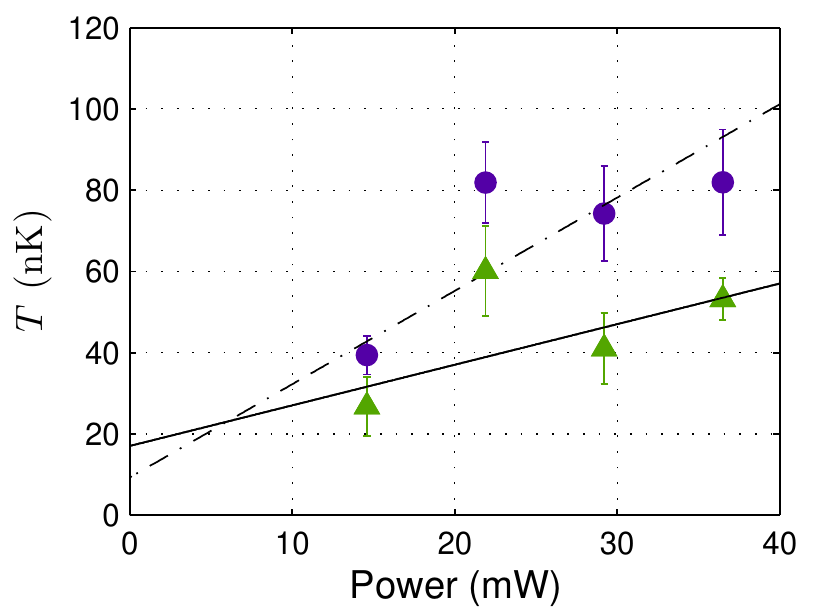}  
\caption{\label{fig:greensheettemp} Measured temperature vs mask beam power. Measurements are for a vertical trapping frequency of 520~Hz (circles) and 970~Hz (triangles) as a function of power in the radial trapping beam incident on the DMD. The experiment was carried out with the lowest radial trapping power of 14.6~mW.}
\end{figure}

A modified procedure is used when the blue detuned vertical trap is used. The blue-detuned vertical trap is a TEM$_{01}$ beam. Atoms initially reside in the combination of red-detuned vertical trap and the crossed dipole trap. The atoms are then adiabatically transferred from the red-detuned vertical trap to the blue-detuned vertical trap (while horizontal confinement is maintained by the crossed dipole trap). We then perform a forced evaporation ramp by lowering the trapping power of the crossed dipole trap. Finally, the atoms are transferred to the target potential and the crossed dipole beam is turned off. We let the  condensate equilibriate for 1~s. The temperature in the blue-detuned trap is set by both the depth of the target trap potential and the power of the blue-detuned vertical trap. The method used to extract temperature from the red-detuned trap does not work with the blue-detuned trap due to the lower temperature. To circumvent this problem, we blow away the atoms in the ring and let the atoms in the disc expand in time of flight, imaging vertically. This is done for two primary reasons. First, the central disc is hard-walled and we expect the atoms in the disc to have a lower critical temperature~\footnote{A simple assumption of uniform density for the hard walled disc puts the critical temperature to be on the order of 100~nK below the critical temperature of atoms in the ring}. A lower critical temperature results in a higher fraction of thermal atoms, making it easier to extract a temperature. Second, an analytical expression for an expanding toroidal trap does not exist~\footnote{we assume that the temperature of atoms in the disc and the ring are equal}. To make our measurements more accurate, we not only took data in the experimental configuration (with a target trap power of 14.6~mW), but also at higher powers using the same atom number and vertical trapping frequency of the blue-detuned trap. A fit of the temperatures measured at higher power can be linearly extrapolated to verify the measured temperature at the experimental configuration. The measured temperatures for the blue-detuned trap are shown in Fig.~\ref{fig:greensheettemp}. We reach a temperature of 40(12)~nK and 30(10)~nK for vertical trap frequencies of 520~Hz and 970~Hz respectively.

\subsection{The effect of temperature on perturbation strength calibration}

\begin{figure}
\centering
\includegraphics[width=80mm]{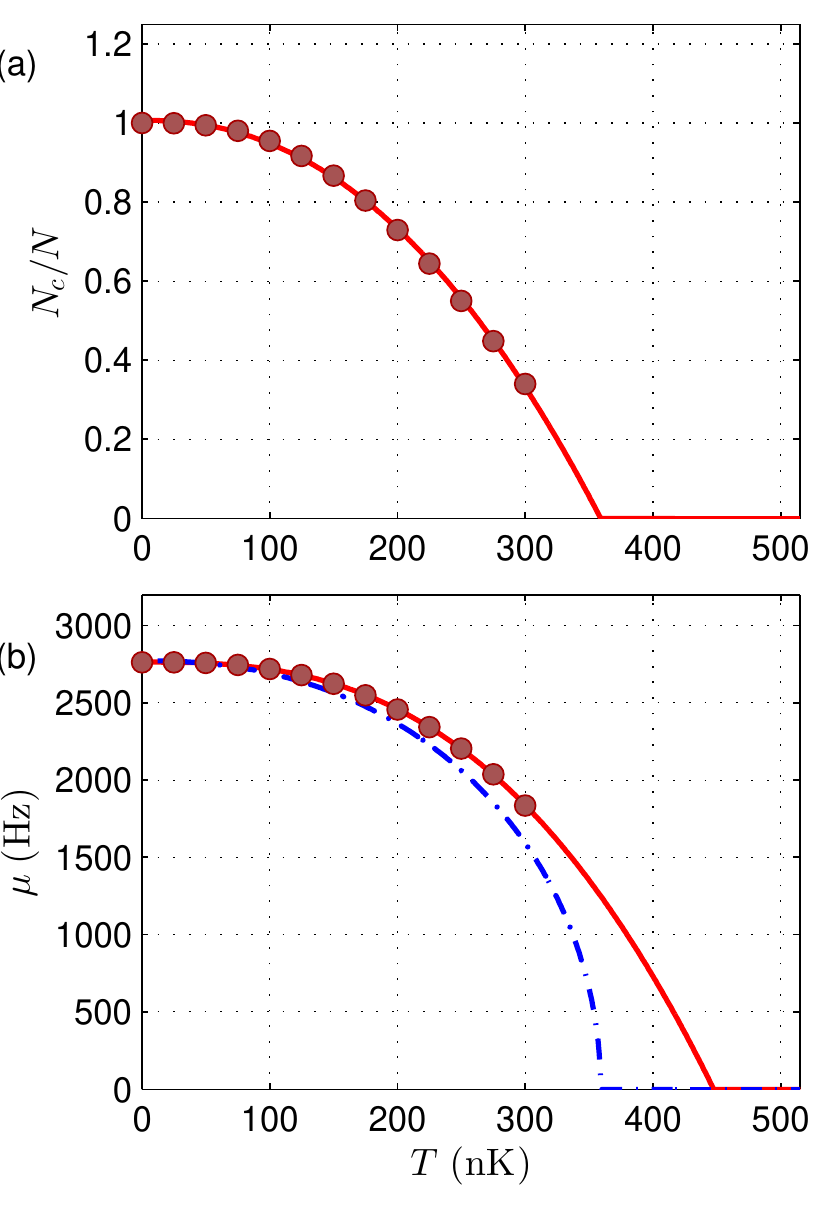}  
\caption{\label{fig:muvstemp} (a) Computed condensate fraction as a function of temperature. The points show the results of our ZNG calculations for a vertical trapping frequency of 512~Hz (see text). A fit of the form $N_c/N = 1 - (T/T_{c})^ \alpha$ with $\alpha=2.22(9)$ is also shown. (b) Computed chemical potential as a function of temperature. The red line shows a fit of the form $\mu(T)/\mu(T=0) = 1 - (T/T_{c})^ \beta$ with $\beta=2.72(4)$. For reference, the dash-dot line shows $\mu(T)/\mu(T=0) = [1 - (T/T_{c})^ \alpha]^{1/2}$, expected from the Thomas-Fermi approximation. The difference between these two curves yields the first order correction to the barrier calibration as a function of temperature.}
\end{figure}

Calibration of the perturbation strength is done {\it in-situ} and follows the same procedure as ~\cite{eckel2014h1}. Briefly, the optical density of atoms at the position of the perturbation is measured as a function of perturbation strength. Due to optical aberrations in the imaging system, the behavior of the optical density vs. $U_b$ changes between $U_b/\mu<1$ and $U_b/\mu>1$. In particular, this function exhibits an ``elbow'' at $U_b/\mu=1$. The location of the elbow where the optical density levels outs enables us to determine the chemical potential of the un-perturbed toroid.

During imaging, we are unable to distinguish thermal atoms from the condensate atoms. It is possible that as we change the temperature, the resulting change in the thermal fraction may impact the measurement of the perturbation strength. Here, we investigate the systematic error introduced due to the barrier calibrations done at different temperatures. We performed ZNG~\cite{zaremba19991} calculations to determine the effect of finite temperature on our measurements. In the ZNG model, the effective potential experienced by the condensate is:
\begin{equation}
U_c=V_{3d}+2gn_{3dt}. 
\end{equation}
Here $V_{3d}$ is the toroidal potential, $n_{3dt}$ is the number density of the thermal cloud, $g = 4\pi \hbar^2 a_{s}/m$ is the interaction strength coefficient and $a_{s}$ is the s-wave scattering length. This enables us to calculate the total number of atoms in the condensate $N_c$ and the density of condensate atoms $n_{3dc}$ by using the Thomas-Fermi approximation. The effective potential felt by the thermal atoms is given by:
\begin{equation}
U_t=V_{3d}+2gn_{3dt}+2gn_{3dc}. 
\end{equation}
This potential $U_t$ is used to determine the thermal atom distribution $n_{3dt}$,
\begin{equation}
n_{3dt} = 1/\Lambda_{dB}^3  \text{Li}_{3/2}(\exp((\mu-U_{t})/k_{B}T)).
\end{equation}
which can be summed up to yield the total number of thermal atoms $N_t$. Here $\text{Li}_{3/2}$ is the polylogarithmic function of order 3/2.  These equations are solved under the constraint that the total atom number $N_0$ is the sum of condensate atom number $N_c$ and thermal atom number $N_t$, and remains constant. For a given temperature, this procedure of calculating the number of thermal atoms $N_t$ and condensate atoms $N_c$ is carried iteratively until the solution converges. The lowest temperature where the condensate atom number drops to zero is the critical temperature $T_c$.

Figure \ref{fig:muvstemp}(a) shows the calculated condensate fraction as a function of temperature for a vertical trapping frequency $\omega_{z}$ of 518(4)~Hz and radial trapping frequency of 258(12)~Hz. The solid line shows a fit of the form $N_c/N = 1 - (T/T_{c})^{\alpha}$ with $\alpha=2.22(9)$. The extrapolated fit yields a critical temperature of 370~nK. Figure \ref{fig:muvstemp}(b) shows the calculated chemical potential as a function of temperature. A fit of the form $\mu(T)/\mu(T=0) = 1 - (T/T_{c})^ \beta$ with $\beta=2.72(4)$ is shown as a solid line. For reference, the dash-dot line shows  the expected Thomas-Fermi chemical potential $\mu(T)/\mu(T=0) = [1 - (T/T_{c})^ \alpha]^{1/2}$. (For a ring, $N\propto \mu^2$ in the Thomas-Fermi approximation.) The shift between these two curves arises from the additional mean-field interaction between the thermal gas and the condensate. For a vertical trapping frequency of 512~Hz, the highest temperature that we operate at is 85(20)~nK, which should be compared to the critical temperature of 370~nK (see Fig. \ref{fig:muvstemp}). The fractional change in chemical potential due to the thermal component is $3.5\times10^{-2}$. This leads to a 3~\% systematic shift in the barrier calibration. At the higher temperature of 195(30)~nK with $\omega_z=985$~Hz and $\omega_r=258(12)$~Hz, the systematic shift is around 8~\%(owing to the higher transition temperature of 470~nK), but this is small compared to the statistical error.

\onecolumngrid 
\section{Table of Experimental parameters and fit}

\begin{table}[h]
\center
\begin{tabular}{lccccccc}
Case   & $T$ (nK) & $T_{c}$ (nK) & $\omega_{z}/2 \pi$~(Hz)   & $N/10^5$    & $\mu/h$ (kHz) & $T_{\rm esc}$~(nK)  &  $\Omega_a$~(s$^{-1}$)    \\
\hline
I   & 30(10)  & 470(30)   & 974(7)  & $4.46(26)$ & 2.91(12) & 3.9(6)$\times10^3$  &   $5(2)\times10^{-1}$  \\
II  & 40(12)  & 370(40)   & 518(4)  & $6.71(39)$ & 2.93(11) & 9.2(8)$\times10^3$  &   $4.8(9)\times10^{0}$   \\
III & 85(20)  & 370(40)   & 520(10) & $6.48(46)$ & 2.68(11) & 5.9(8)$\times10^3$  &   $1.9(4)\times10^{3}$  \\
IV  & 195(30) & 470(30)   & 985(4)  & $4.22(26)$ & 2.66(08) & 3.2(4)$\times10^3$  &   $1.2(2)\times10^{5}$  \\
\hline\hline
\end{tabular}
\caption{\label{tab:config}  The temperature ($T$), critical temperature $T_{c}$, vertical trapping frequency $\omega_{z}$, number of atoms $N$, chemical potential ($\mu$) and fit parameters escape temperature $T_{\rm esc}$ and $a$ for different trapping configurations. The radial trapping frequency $\omega_{r}$ remains essentially constant across all the configurations at 258(12)~Hz. Errorbars in $N$ and $\mu$ exclude systematic effects which we estimate to be up to a 20\% common shift.}
\end{table} 
\twocolumngrid

\end{document}